\documentclass[12pt]{article}

\usepackage[margin=1.3in]{geometry}
\usepackage{graphicx}
\usepackage{hyperref}
\usepackage{amsfonts}
\usepackage{amsthm}
\usepackage{amsmath}
\usepackage{float}

\usepackage{babel}
\usepackage{geometry}
\usepackage{titling}
\usepackage{blindtext}

\title{Optimizing Sparse Mean-Reverting Portfolio}
\author{Sung Min Yoon}
\date{sy375@cornell.edu} 

\begin{document}

\maketitle
\begin{sloppypar}

\vspace{1in}

\begin{abstract}

Mean-reverting behavior of individuals assets is widely known in financial markets. In fact, we can construct a portfolio that has mean-reverting behavior and use it in trading strategies to extract profits. In this paper, we show that we are able to find the optimal weights of stocks to construct portfolio that has the fastest mean-reverting behavior. We further add minimum variance and sparsity constraints to the optimization problem and transform into Semidefinite Programming (SDP) problem to find the optimal weights. Using the optimal weights, we empirically compare the performance of contrarian strategies between non-sparse mean-reverting portfolio and sparse mean-reverting portfolio to argue that the latter provides higher returns when we take into account of transaction costs.
  
\end{abstract}

\newpage

\section{Introduction}
The idea of mean-reversion has been around in financial markets as an indicator of predictability in returns over long horizons. In fact, in generating profits through trading, it is crucial for the trader to correctly choose between momentum and mean-reversion strategies. In this paper, we focus on the idea of mean-reversion and construct mean-reverting portfolios mainly by solving optimization problems. We introduce the semidefinite programming (SDP) problems in building mean-reverting portfolios with additional conditions such as sparsity and apply contrarian strategy to see whether the optimal portfolios are able to produce positive returns. This paper closely follows the work by Cuturi (2015) and d'Aspremont (2011).

While the definitions of momentum and mean-reversion may be slightly different based on how a person uses and which horizon a person looks at, momentum strategies assume that the market will continue to move in the same direction and mean-reversion strategies assume that the market will partly reverse back. There has been many literature and trading strategies that were based on momentum of stocks (i.e., buying the winners and selling the losers), and they are quite common strategies in financial markets today. There has also been quite a few studies on mean-reversion in univarite time series (single stock). However, due to the complexity in problems, the idea of mean-reverting portfolios did not get much attention. It is also hard to extract meaningful profits from mean-reverting portfolios because they often behave like noise (i.e., they have small variance) and require large transaction costs compared to using the idea of mean-reversion in individual stocks.

In order to make mean-reverting portfolios more attractive to trading, we can add additional constraints to the optimal portfolio. Here, we define mean-reverting portfolio as those having fast mean-reversion character based on predictability estimator, which will be discussed in detail in Section 2. The optimal mean-reverting portfolios would have constant mean and variance and tend to revert back to its mean faster than other portfolios with the same asset classes and different weights. We look at two additional factors to extract large profits from using contrarian strategy, which is basically buying the portfolio when the portfolio is below the mean-reverting line and selling the portfolio when the portfolio is above the mean-reverting line. First, it is important that the mean-reverting portfolio has large enough variance. The contrarian strategy basically extracts the difference between the portfolio's position and its mean-reverting position. Thus, the trader would be able to make more profits by using contrarian strategy to the mean-reverting portfolio with large enough variance compared to the one with very small variance (e.g., portfolio that stays constant over time). Second, it is important that the mean-reverting portfolio is sparse. If we include too many stocks in our mean-reverting portfolio, the transaction costs would outweigh the profits from contrarian strategies.

In finding the optimal weights for portfolio with the fastest mean-reversion, we can actually use the idea of canonical correlation analysis to achieve closed-form solution (Section 3.1). However, if we include constraints on minimum variance and sparsity of the portfolio, we can only find the optimal weights by solving optimization problem. Since the optimization problems to find optimal weights directly are hard to solve, we transform the problem into semidefinite programming (SDP) problem by convex relaxation. For variance, we can directly include the constraint on minimum variance in the SDP problem. However, for sparsity, it is hard to enforce limit on the number of non-zero elements in the optimal weights. Hence, we instead include regularization factor into the objective function to enforce sparsity.

In the last part of the paper, we actually apply contrarian strategy to the optimal mean-reverting portfolios to the implied volatility data of S\&P 500 stocks. We compare the returns of mean-reverting portfolio with constraints on minimum variance and sparsity with the ones without constraints and argue that adding constraints improves the performance when applied to trading after accounting for transaction costs.

\section{Theoretical Framework}
Throughout the paper, we want to find a basket of portfolio that has mean-reverting behavior with additional constraints that will be described in the following sections. To analyze portfolio over time, we need to look into a vector that consists of $n$ values, which depicts the number of stocks or instruments in the portfolio, for $t$ time periods, and the weights of each stock or instrument. Thus, we formulate a problem into looking at a vector valued process $x=(x_t)_{t\in \mathbb{N}}$ where each $x_i \in \mathbb{R}^n$ and finding the optimal vector $y \in \mathbb{R}^n$. In this paper, we want the process $(y^Tx_t)$ to be a stationary process, which is one of the most popular processes that has mean-reverting behavior.

In analyzing the portfolio, we need to use the autocovariance matrix of $x_t$. We define lag-k autocovariance matrix of $x_t$ as follows:
\begin{equation}
    A_k = \frac{1}{T-k-1}\displaystyle\sum_{t=1}^{T-k}\tilde{x}_t\tilde{x}_{t+k}^T, \tilde{x}_t=x_t-\frac{1}{T}\displaystyle\sum_{t=1}^{T}x_t
\end{equation}
for $x=(x_1,...,x_T)$ and $k\ge0$.

For such a process, we first apply canonical decomposition derived in Box and Tiao (1977):
\begin{equation}
    x_t=\hat{x}_{t-1}+\epsilon_t
\end{equation}
where $\hat{x}_{t-1}$ is a predictor of $x_t$ predicted through the past values of the process $(x_1,...,x_{t-1})$ and $\epsilon_t$ is a vector of independent and identically distributed Gaussian noise with zero mean and covariance matrix $\Sigma$.

Using this canonical form of the process and assuming that the expectation of the process is 0 (this can be achieved through subtracting the mean for each value $x_t$ and $\hat{x}_{t-1}$), we have:
\begin{equation}
    \mathbb{E}[x_t^2]=\mathbb{E}[\hat{x}_{t-1}^2]+\mathbb{E}[\epsilon_t^2]
\end{equation}
for univariate case.

\subsection{Estimators}
For a stationary process, there are several estimators that measure the degree of mean-reversion: Predictability (Box and Tiao, 1977), Portmanteau statistics (Ljung and Box, 1978), and Crossing Statistic (Kedem and Yakowitz, 1994). In all three estimators, a process having a low value implies that the process is fastly mean-reverting. While any and all of the estimators can be used in formulating optimization problems, we focus on predictability estimator, which is the most popular estimator for stationary processes, in our optimization problems.

\subsubsection{Predictability}
Box and Tiao (1977) have defined the predictability of $x_t$ by the ratio of the variances of $x_t$ and $\hat{x}_{t-1}$. We let $\sigma^2$ and $\hat{\sigma}^2$ be the variances of $x_t$ and $\hat{x}_{t-1}$ respectively and define the predictability $\lambda$ of $x_t$ as:
\begin{equation}
    \lambda=\frac{\hat{\sigma}^2}{\sigma^2}
\end{equation}

From (3), we have $\sigma^2=\hat{\sigma}^2+\Sigma$, which gives $\lambda = \frac{\hat{\sigma}^2}{\sigma^2}=1-\frac{\Sigma}{\sigma^2}$. Thus, the smaller the $\lambda$ is, the smaller the variance of the noise dominates the variance of $x_{t-1}$, and $x_t$ is almost pure noise ofr very small $\lambda$. In contrast, for larger $\lambda$, $\hat{x}_{t-1}$ dominates the noise and $x_t$ can almost perfectly be predicted.

In multivariate process $(y^Tx_t)_t$, we have $y^Tx_t=y^T\hat{x}_{t-1}+y^T\epsilon_t$ from (2) and the predictability of this is defined as:
\begin{equation}
    \lambda(y)=\frac{y^T\hat{A}_0y}{y^TA_0y}
\end{equation}
where $A_0$ and $\hat{A}_0$ are covariance matrices of $x_t$ and $\hat{x}_{t-1}$ respectively in a way that can be calculated by (1).




\section{Finding Optimal Baskets}
\subsection{Minimizing Predictability Without Constraints}
Since we want our mean-reverting portfolio to have large mean-reverting behavior and be unpredictable, we would want to minimize the predictability. To minimize the predictability $\lambda(y)$ of the portfolio (multivariate case), we have to find the minimum generalized eigenvalue $\lambda$ that solves $\det(\lambda A_0 -\hat{A}_0)=0$.

Since $A_0$ is a covariance matrix, it is a positive semi-definite matrix. However, if we further assume that $A_0$ is positive definite, the minimum predictability of the multivariate process $(y^Tx_t)_t$ is:
\begin{equation}
    y=A_0^{-1/2}y_0
\end{equation}
where $y_0$ is the eigenvector corresponding to the smallest eigenvalue of $A_0^{-1/2}\hat{A}_0A_0^{-1/2}$.

While we have this closed form solution to the minimization problem, we cannot simply calculate it because $\hat{A}_0$ is based on the predictor $\hat{x}_{t-1}$. Thus, we need an estimate of the matrix $\hat{A}_0$. Box and Tiao (1977) have come up with an estimation of $\lambda(y)$ using the autoregression model of order $p$ (VAR($p$)) to come up with $\hat{\lambda}(y)$ by using $A_1A_0^{-1}A_1^T$ and an estimate for $\hat{A}_0$. Therefore, our goal would be:
\begin{equation}
    \min_y\hat{\lambda}(y) \textrm{ where } \hat{\lambda}(y)=\frac{y^T(A_1A_0^{-1}A_1^T)y}{y^TA_0y}
\end{equation}
This implies that the optimal $y$ would be the eigenvector that corresponds to the smallest eigenvalue of the matrix $A_0^{-1/2}A_1A_0^{-1}A_1^TA_0^{-1/2}$ (assuming $A_0$ is invertible), which means that we have a closed form solution to our minimization problem. We can use this solution as the optimal baskets of the portfolio that has mean-reverting behavior and minimum predictability.
\begin{figure}[h!]
\begin{center}
\includegraphics[scale=0.4]{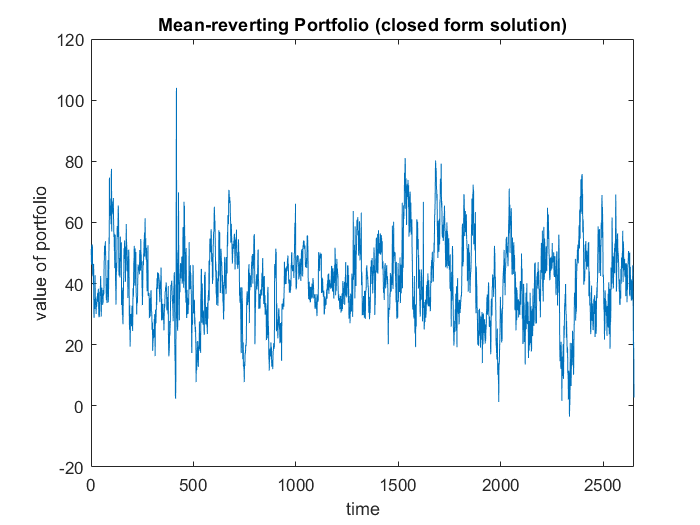}
\includegraphics[scale=0.4]{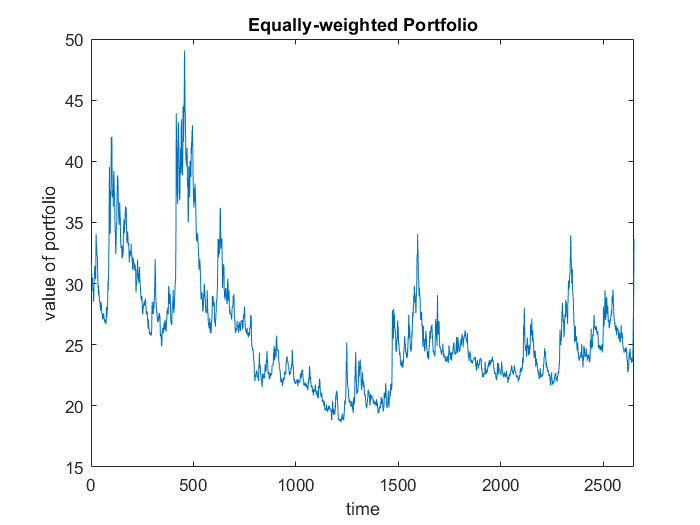}
\end{center}
\caption{Example of mean-reverting portfolio using optimal weights derived from the closed-form solution (left) and equally-weighted portfolio (right)}
\label{setup}
\end{figure}

Here, we can clearly see that we are able to produce a mean-reverting portfolio using the weights from the closed-form solution stated above. The optimal mean-reverting portfolio has constant mean and variance and reverts back to its mean fast, especially comparing it with a portfolio that has equal weights of all the individual stocks.

\subsection{Minimizing Predictability With Variance Constraints}
To use mean-reverting portfolio in contrarian strategies, we want our portfolio to have maximal mean reversion (minimum predictability) while having sufficiently large volatility (variance). One way to achieve this is to enforce the portfolio to have some minimum level of variance. In this case, we have the following optimization problem:
\begin{equation}
\begin{array}{rrclcl}
\displaystyle \min_{y} & \multicolumn{3}{l}{y^TMy} \\
\textrm{s.t.} & y^TA_0y & \geq & \nu \\
& ||y||_2 & = & 1 \\
\end{array}
\end{equation}
where $M=A_1A_0^{-1}A_1^T$ and $\nu$ is the minimum variance we want to achieve.

In our problem, we also want to study the optimal weights after enforcing sparsity. In other words, we want to solve the optimization problem in (8) with additional constraint regarding the number of non-zero entries in $y$. One way to do this is to limit the number of non-zero entries to $k$. In fact, the problem would be the same as to setting the 0-norm equal to $k$ instead of smaller than equal to $k$:
\begin{equation}
\begin{array}{rrclcl}
\displaystyle \min_{y} & \multicolumn{3}{l}{y^TMy} \\
\textrm{s.t.} & y^TA_0y & \geq & \nu \\
& ||y||_2 & = & 1 \\
& ||y||_0 & = & k \\
\end{array}
\end{equation}

\section{Semidefinite Relaxation and Sparsity Constraints}
We would now like to solve the optimization problem that we formulated  in (8) or (9). However, these problems are often extremely hard to solve as they are not convex and involve sparse selection of variables. One way to solve this issue is to make some transformation to the equation and relax some constraints. Through such process, we end up with Semidefinite Programming (SDP) problems, which are relatively easy to solve. We now show how we transform the problems into SDP problems.

We try to formulate the problem using $Y=yy^T$ instead of $y$. Since $Y$ has only positive semidefinite entries, we know that $Y$ is a semidefinite variable. We can easily show that $y^TMy=Tr(Myy^T)=Tr(MY)$ through expansion. Likewise, we have $y^TA_0y=Tr(A_0Y)$. The constraint on the square of individual components of $y$ ($||y||_2=1$) can be translated into $Tr(Y)=1$. Also, we need additional constraint on the rank of $Y$ and the eigenvalues of $Y$ for it to satisfy that it can be written as $Y=yy^T$ and it is a semidefinite matrix. Lastly, we change our sparsity constraint where we directly enforced the number of entries to $k$ into adding regularization factor to the objective function. While we are not able to maintain direct relationship between $k$ and our problem, we can find the regularization factor that gives $k$ number of non-zero entries of $y^*$ empirically by trying multiple values for the regularization factor $\rho$. Incorporating all these, we end up with the following SDP problem:
\begin{equation}
\begin{array}{rrclcl}
\displaystyle \min_{Y} & \multicolumn{3}{l}{Tr(MY)+\rho ||Y||_1} \\
\textrm{s.t.} & Tr(A_0Y) & \geq & \nu \\
& Tr(Y) & = & 1 \\
& Rank(Y) & = & 1 \\
& Y & \succeq & 0
\end{array}
\end{equation}
where we define $||Y||_1=\displaystyle\sum_{i, j} |Y_{ij}|$ and $\rho > 0$ is a regularization factor. Here, $Y \succeq 0$ means that the smallest eigenvalue is nonnegative (this is a common notation in literature that discusses SDP). 

While we have achieved our goal of transforming the problem into SDP problem, we still cannot easily solve (10). This is mainly due to the constraint on the rank of $Y$. Thus, we consider relaxing this problem by dropping the rank constraint:
\begin{equation}
\begin{array}{rrclcl}
\displaystyle \min_{Y} & \multicolumn{3}{l}{Tr(MY)+\rho ||Y||_1} \\
\textrm{s.t.} & Tr(A_0Y) & \geq & \nu \\
& Tr(Y) & = & 1 \\
& Y & \succeq & 0
\end{array}
\end{equation}
Brickman (1961) actually showed that when there is no regularization (i.e., when $\rho=0$), solving this problem (11) actually produces the same solution to the one in (10). This means that (11) automatically produces $Y$ that has rank 1 when there is no regularization.

In order to recover back the optimal solution $y^*$ from $Y^*$, we extract the largest eigenvalue of $Y^*$ and find the corresponding eigenvector. When $\rho=0$, we can in fact use the corresponding eigenvector of the largest eigenvalue as the optimal basket $y^*$ of the portfolio since $Y^*$ is a rank 1 matrix. When $\rho>0$, we are not able to maintain rank 1 condition of $Y^*$. However, when $\rho$ is sufficiently small enough, we obtain 1 eigenvalue that is large and $n-1$ eigenvalues that are very close to 0. Thus, we use the same technique of using the largest eigenvector to predict $y^*$ from $Y^*$.

\begin{figure}[h!]
\begin{center}
\includegraphics[scale=0.4]{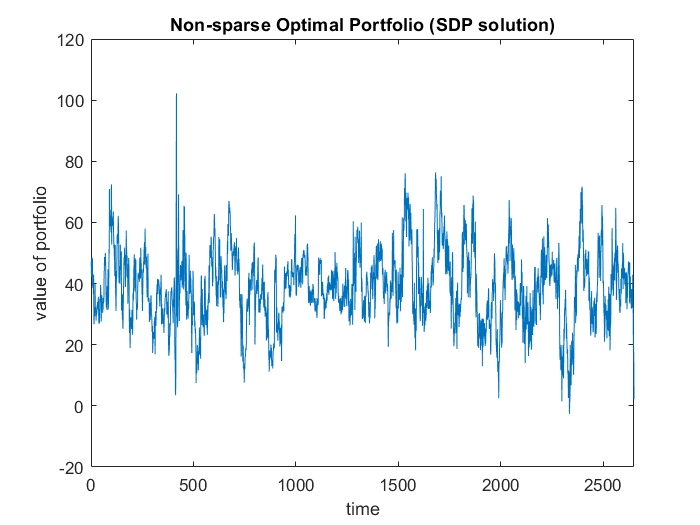}
\includegraphics[scale=0.4]{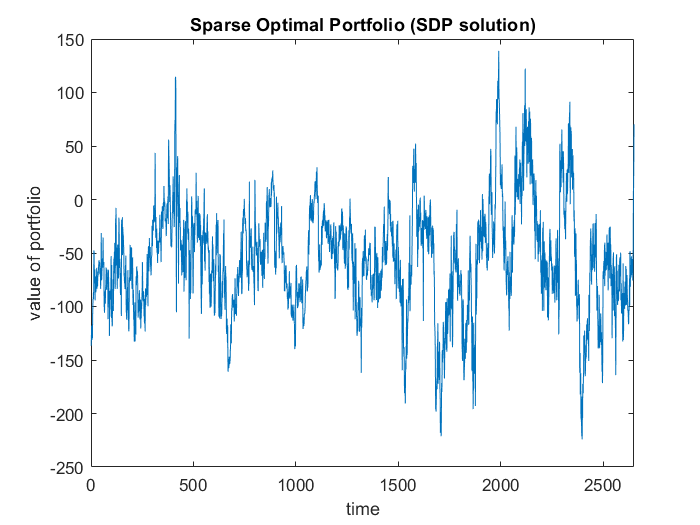}
\end{center}
\caption{Example of non-sparse ($\rho=0$) mean-reverting portfolio using optimal weights derived from the optimization solution (left) and sparse ($\rho=0.2$) mean-reverting portfolio with minimum variance requirement ($\nu=0.1$) using optimal weights derived from the optimization solution (right)}
\label{setup}
\end{figure}

Here, by comparing with the left figures in Figure 1 and Figure 2, we can see that the optimal portfolio using the weights from solving SDP problem without constraints ($\rho=0, \nu=0$) produces a very similar, almost identical outcome to the optimal portfolio using the closed form solution.

If we compare the two figures in Figure 2, we can see that while non-sparse optimal portfolio (left) has faster mean-reverting behavior, we have achieved our goal of enforcing sparsity and minimum variance requirement with mean-reverting behavior (right).

\section{Application to Trading}
We now want to apply our optimal baskets of portfolio to trading and see how the profits look. The trading strategy we will use is a contrarian strategy (more specifically, Jurek and Yang Trading Strategy). Since we know that our optimal portfolio has mean-reverting behavior and has large variance, we can extract the profit by selling the winners and buying the losers as we know that the value of the portfolio will revert back to its mean.

Using such strategy, we first show that our optimal baskets of portfolio are able to generate profits. We then incorporate the effect of transaction costs and compare the profits between optimal baskets with and without constraints.

\subsection{Trading Strategy}
Among various contrarian strategies, the strategy that we chose to use is Jurek and Yang (2007) trading strategy. We first find the optimal weights of the portfolio using the techniques in part 4 and treat this portfolio as one instrument. Because we know that this portfolio has mean-reverting behavior, whenever the portfolio's position is above the expected mean, we sell the portfolio proportional to the difference in the position and the expected mean. On the other hand, whenever the portfolio's position is below the expected mean, we buy the portfolio proportional to the difference in the position and the expected mean.

Jurek and Yang used the above logic to construct a trading strategy. They proposed the trading strategy that for stationary process $P_t$ that is governed with $P_{t+1}=\lambda P_t +\sigma \epsilon_t$ with mean $\bar{P}$, one can take contrarian position by taking a position $N_t$ (number of units of portfolio) proportional to the following:
\begin{equation}
    N_t = \frac{\lambda(\bar{P}-P_t)}{\sigma^2}W_t
\end{equation}
where $W_t$ is the investor's wealth at time $t$. This is clearly a contrarian strategy since whenever $P_t>\bar{P}$, we are enforcing negative position (short) of the portfolio and whenever $P_t<\bar{P}$, we are enforcing positive position (long) of the portfolio.

Here, however, we are not able to get $\lambda$ and $\sigma$ for real instruments. Thus, we use the estimate of $\lambda$ and $\sigma$ that we find through empirical research. We use the estimators that d'Asprement (2011) proposed to get the estimates, which are:

\begin{align}
\hat{\mu} &= \frac{1}{T} \displaystyle\sum_{t=1}^T P_t \\
\hat{\lambda} &= -\log\left(\frac{\displaystyle\sum_{t=1}^T (P_t-\hat{\mu})(P_{t-1}-\hat{\mu})}{\displaystyle\sum_{t=1}^T (P_t-\hat{\mu})(P_{t}-\hat{\mu})}\right) \\
\hat{\sigma}^2 &= \frac{2\lambda}{(1-e^{-2\lambda})(N-2)}\displaystyle\sum_{t=1}^T ((P_t-\hat{\mu})-e^{-\lambda}(P_{t-1}-\hat{\mu}))^2
\end{align}
\\
\\
These can all be calculated with our optimal portfolio by looking at the historical positions of the portfolio $P_1, P_2, ..., P_T$. We can now treat $\hat{\lambda}$ and $\hat{\sigma}$ as fixed values.

We can then use these estimates to construct a position of the portfolio. At certain time $t$, we want to have $N_t = \frac{\hat{\lambda}(\bar{P}-P_t)}{\hat{\sigma}^2}W_t$ number of units of portfolio. Then, at (the beginning of) time $t+1$, we would have wealth of $W_{t+1}=W_t+N_t(P_{t+1}-P_t)$ from holding $N_t$ number of units of portfolio. Then, we rebalance our position to $N_{t+1} = \frac{\hat{\lambda}(\bar{P}-P_{t+1})}{\hat{\sigma}^2}W_{t+1}$ using the new wealth level $W_{t+1}$.

\subsection{Experiment Setup}
To see how our optimal portfolio and trading strategy actually works, we experiment on the daily time series of option implied volatilities of S\&P 500 stocks from January 4, 2010 to February 28, 2020. After removing stocks that do not have any of the values in the period, we are left with 448 stocks. The key advantage of using the data on option implied volatility is that these numbers tend to vary in a quite limited range. Also, volatility tend to be stationary, which makes it helpful in our analysis. While option volatility is not directly tradable, we assume that it can be synthesized using baskets of call options and assimilate it to tradable asset.

\subsubsection{Data Selection}
Among 448 stocks, we randomly selected 12 stocks to test our trading strategy. The selected stocks were: Aflac Incorporated (AFL), Bank of America (BAC), BorgWarner (BWA), Eastman (EMN), Kellogg (K), Kohl's (KSS), Loews Corporation (L), Norfolk Southern Corp. (NSC), Paychex (PAYX), PNC Financial (PNC), Starbucks (SBUX), and Western Digital (WDC). It turns out that our random selection of stocks does not rely on specific sector.

The next step to our setup would be to divide the data into training and testing sets. We use the data from 2010 to 2018 for training (i.e., to use in finding the optimal weights and estimates for $\lambda$ and $\sigma^2$) and data from 2019 to 2020 for testing. Since we only have first 2 months of data in 2020, we end up with roughly 9:1 split for training and testing data. For testing data, we are left with 289 trading days, on which we rebalance our position.

\subsubsection{Transaction Costs}
For transaction costs, we assume that fixed transaction costs are negligible. However, we consider that there exist transaction costs that are incurred each trading date and are proportional to the change in position and number of assets in the portfolio. We vary this transaction cost to be 0.04 cents, 0.08 cents, 0.12 cents, and 0.16 cents per contract for robustness check.

In more detail, in each trading date, we assume the total transaction cost to be $tc*k*|N_{t}-N_{t-1}|$ where $tc$ is the transaction cost ($tc \in \{0.0004, 0.0008, 0.0012, 0.0016\}$), $k$ is the number of assets in the portfolio, and $|N_{t}-N_{t-1}|$ is the net change in position of the portfolio. We subtract this value to calculate the wealth at each period.

\subsection{Empirical Results}
We now apply our optimal weights and trading strategy to compare the performance.

For a sparse portfolio, we use $\rho=0.2$ for regularization factor and $\nu=0.1$ for the minimum variance requirement. With these constraints, we end up with an optimal basket that uses 5 stocks (out of 12 stocks).
\begin{figure}[H]
\begin{center}
\includegraphics[scale=0.5]{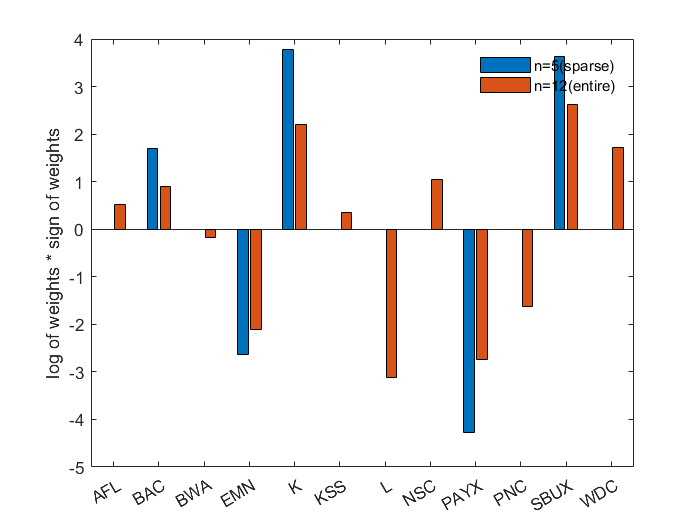}
\end{center}
\caption{Individual weights of 12 stocks for sparse portfolio (blue) and non-sparse portfolio (red). We display log of weights multiplied by the sign of weights for visualization purpose. Here, we can see that the sparse portfolio is constructed of 5 stocks and non-sparse portfolio is constructed of 12 stocks.}
\label{setup}
\end{figure}

We now use these weights to construct optimal portfolios and assume it as a single instrument. We first consider the case where there is no transaction cost (i.e., $tc=0$).
\begin{figure}[H]
\begin{center}
\includegraphics[scale=0.5]{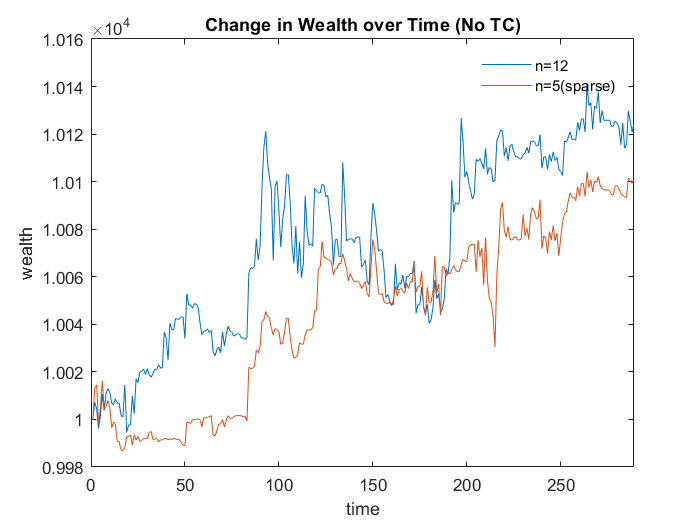}
\end{center}
\caption{Change in wealth over time when there is no transaction costs}
\label{setup}
\end{figure}

In Figure 4, we can clearly see that the optimal portfolio without constraints outperforms the optimal portfolio with constraint. The optimal portfolio without sparsity and variance constraint has generated 1.2\% return over the 14-month period, which is 0.2\% larger than the return from sparse optimal portfolio. This is because the one that uses the entire constraint is likely to have faster mean-reversion character, which will allow the investor to extract larger profits through using the Jurek and Yang trading strategy.

In practice, there exists transaction costs when we execute trades. Thus, we now take into consideration of the transaction costs in trading. We use the same optimal portfolios for the one with and without sparsity constraint and use Jurek and Yang trading strategy to various levels of transaction cost.

\begin{figure}[h!]
\begin{center}
\includegraphics[scale=0.4]{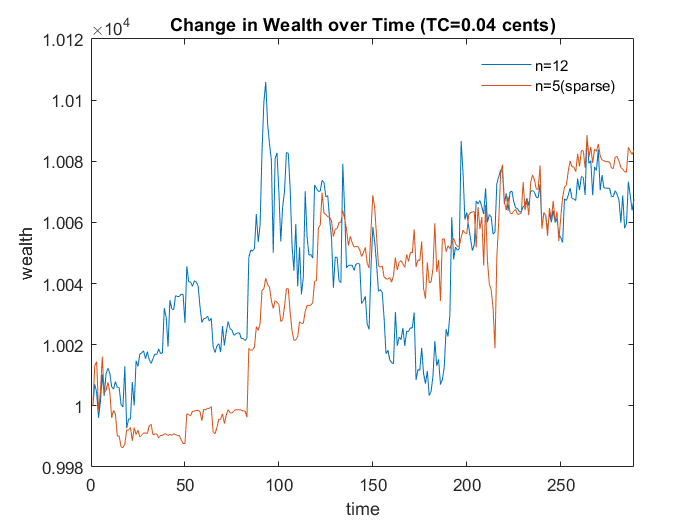}
\includegraphics[scale=0.4]{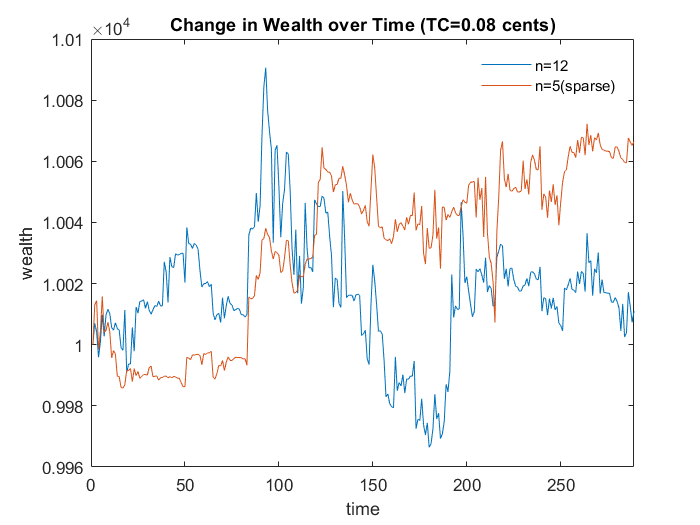}
\includegraphics[scale=0.4]{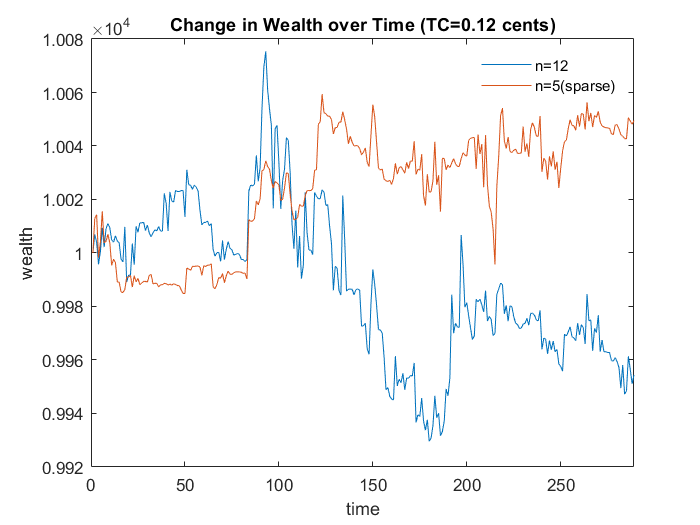}
\includegraphics[scale=0.4]{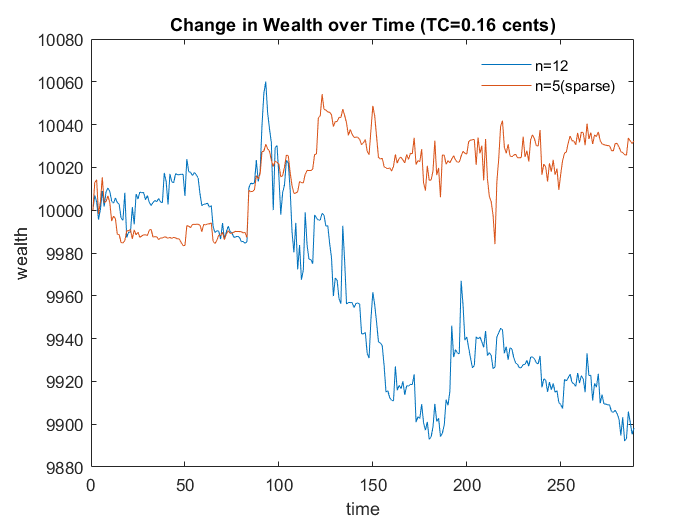}
\end{center}
\caption{Change in wealth over time for different transaction costs (0.04 cents, 0.08 cents, 0.12 cents, and 0.16 cents per contract)}
\label{setup}
\end{figure}

According to Figure 5, we can see that the returns from the sparse optimal portfolio outperforms the returns from the non-sparse optimal portfolio. The difference in returns gets larger as the transaction cost rate increases, which implies that it is important to have sparse set of stocks in the portfolio when the transaction cost is high. In fact, when transaction cost that are above 0.10 cents per contract, the payoff from the sparse optimal portfolio still remains positive while the payoffs from the non-sparse optimal portfolio is negative. This means that the transaction costs outweigh the profits one can extract from trading non-sparse optimal portfolio.

\section{Conclusion}
Throughout the paper, we have explored how to minimize predictability to find mean-reverting portfolios. We have added constraints on minimum variance and sparsity of the portfolio and transformed the optimization problem into semidefinite programming (SDP) problem to find the optimal weights of portfolios. Then, we have empirically tested how the optimal portfolios perform by using the option implied volatility data of S\&P 500 from 2010 to 2020. In particular, we have compared the returns of portfolio with and without constraints on minimum variance and sparisty when we apply Jurek and Yang trading strategy. From this, we were able to find that we can extract optimal weights by solving the SDP problems and the sparse portfolio actually outperforms when we take into account of transaction costs. However, the returns were quite low with under 1.2\% with no transaction costs and under 1\% with transaction costs over the 14-month period.

For further development, it would be nice to see how the trading performance changes based on the rebalancing period. In our research, we have rebalanced the holdings of mean-reverting portfolio each trading day. However, mean-reversion might be more effective if implemented in longer rebalancing periods, such as 3-trading days or 1-week, or shorter rebalancing periods (hours or minutes) if we have available data. Also, it would be interesting to study the optimal values for regularization factor and minimum variance to see how they affect the performance. We might be able to find the optimal number of assets in the mean-reverting portfolio for each transaction costs. Lastly, we can study the best selection of stocks in constructing the portfolio. In our research, we have randomly selected 12 stocks among 448 stocks in S\&P 500 to construct initial portfolio and added constraints to select 5 stocks among them. However, the selection of these 12 stocks might not be the most effective choice to construct mean-reverting portfolios. Thus, we can greedily search the best selection of stocks or apply naive Principal Component Analysis (PCA) to select the stocks for the initial mean-reverting portfolio (mean-reverting portfolio without constraints).

\end{sloppypar}
\end{document}